\documentclass[twocolumn,twoside,slac_two]{revtex4}
\pdfoutput=1
\usepackage{graphicx}
\usepackage{fancyhdr}
\usepackage{amssymb}
\usepackage{url}

\begin{document}

\title{First results from the ANTARES neutrino telescope}

\author{Th.~Eberl for the ANTARES collaboration}
\affiliation{ECAP - Erlangen Centre for Astroparticle Physics,
  Erwin-Rommel-Str.~1,  91058 Erlangen , Germany}

\begin{abstract}
The ANTARES detector is the most sensitive neutrino telescope
observing the southern sky and the world's first particle detector operating in the
deep sea. It is installed in the Mediterranean Sea at a depth of $2475\,\mathrm{m}$.
As an example of early results, the determination of the atmospheric muon flux
is discussed and a good agreement with previous measurements is found. Furthermore,
the results of a search for high-energy events in excess of the 
atmospheric neutrino flux are reported and significant limits are set on the diffuse
cosmic neutrino flux in the multi-TeV to PeV energy range. Using data
from more than 800 days of effective data taking, partly during the
construction phase, a first analysis searching for point-like excesses in the neutrino
sky distribution has been performed. The resulting sensitivity of ANTARES is reported and
compared to measurements of other detectors. A method employed for a
first search for neutrinos from Fermi-detected gamma-ray flaring
blazars in the last 4 months of 2008 is described and the results are reported. No significant
neutrino signal in excess of that expected from atmospheric background has been found. 

\end{abstract}

\maketitle

\thispagestyle{fancy}

\section{INTRODUCTION}

The goal of high-energy neutrino astronomy is to provide a new view 
of the Universe by detecting the messengers emitted from its most violent regions.
The emission of high-energy neutrinos necessarily implies the 
presence of highly relativistic baryons at the acceleration sites and consequently 
provides incontrovertible evidence for the acceleration of charged cosmic rays.
The observable neutrino flux is expected to be generated mainly through 
charged pion production in collisions of high-energy protons from the 
cosmic ray accelerators with the ambient gas or with radiation fields \cite{nuprod}.
These neutrinos will point back to even very distant sources, as they 
are neither absorbed nor deflected, a property that makes these particles
unique astronomical messengers \cite{greisen}.
High-energy neutrinos can be of galactic and extragalactic
origin. Supernova remnants and micro-quasars are examples for candidate sources in our
Galaxy, while gamma-ray bursts and active galactic nuclei represent promising 
potential extragalactic sources \cite{particleastrophysics}.
Due to the extremely low cross section of
neutrino interactions, neutrino detectors
need to instrument very large volumes and should be built in a low background environment.
The current neutrino telescopes exploit the idea, proposed by Markov \cite{markov},
of instrumenting
a large volume of water or ice, in order to detect the charged leptons (in particular
muons) emerging from charged-current neutrino-nucleon interactions.
For a list of recent reviews see
\cite{reviews}.

\section{THE ANTARES DETECTOR}
\begin{figure*}[t]
\centering
\includegraphics[width=135mm]{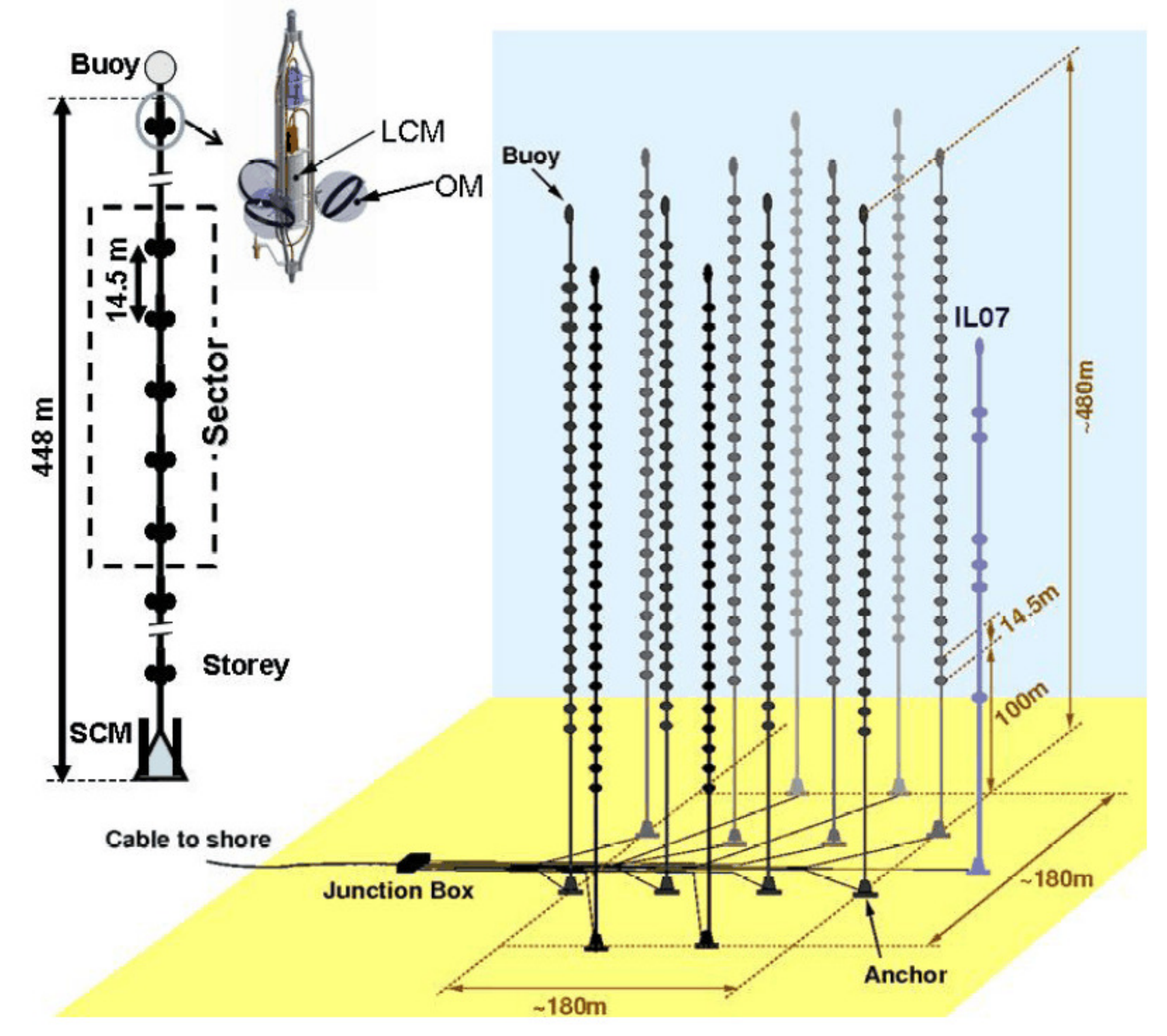}
\caption{Schematic view of the ANTARES detector \cite{detector}.} \label{fig:schematic}
\end{figure*}

The ANTARES detector \cite{detector} (see Fig.~\ref{fig:schematic} for a schematic view)
is located at a depth of $2475\,\mathrm{m}$ in the Mediterranean Sea
($42^{\circ}48'\,\mathrm{N}$, $6^{\circ}10'\,\mathrm{E}$),
$42\,\mathrm{km}$ from the French city of Toulon. It is equipped
with 885 optical sensors arranged on 12 flexible lines. Each line comprises up to
25 detection storeys, each equipped with three downward-looking 10-inch photomultipliers
(PMTs), oriented at 45$^{\circ}$ relative to the vertical. Each PMT is installed in an
Optical Module (OM) that consists of a 17-inch glass sphere in which the optical
connection between the PMT and the glass is assured by an optical gel. Each line is roughly
$450\,\mathrm{m}$ long and is held tight by a buoy at its top. 
The spacing between storeys is $14.5\,\mathrm{m}$.
The distance between adjacent lines is of the order of $60 - 70\,\mathrm{m}$. 
ANTARES contains in addition a line (IL07) with oceanographic sensors
dedicated to the measurement of environmental parameters. ANTARES therefore represents 
an important multidisciplinary 
deep-sea research infrastructure delivering unique data to 
marine biologists and oceanographers.
The construction of the ANTARES detector took place in several sea campaigns
starting in the year 2006 and has been completed in May 2008 with the
deployment and connection of the last 2 lines. Due to this long construction phase and the
modularity of the detector, the commissioning phase comprised several detector 
configurations with different numbers of active lines.
The neutrino detection relies on the emission of Cherenkov photons by high-energy
muons originating from charged-current neutrino-nucleon interactions in or around the 
instrumented volume.
From the PMT positions and the relative arrival times of the 
Cherenkov photons at the PMTs, and making use of the characteristic emission angle 
of Cherenkov radiation, the trajectory of the muon can be reconstructed. 
The direction of
incident neutrinos can be inferred with an energy dependent 
precision that is expected to be better than 0.3 degrees for $E_{\nu}>10\,\mathrm{TeV}$.
As the detector lines move with the deep-sea current, the position of the OMs 
need to be monitored to
ensure this excellent angular resolution. The position of the OMs is
determined every 2 minutes by means of an acoustic triangulation system, while
the orientation of each storey is measured with a compass and a tiltmeter. The timing
calibration \cite{timecal}, which is also crucial for the angular resolution and very stable in time,
is monitored regularly in-situ with dedicated pulsed light sources distributed along the lines.

\section{ATMOSPHERIC MUON FLUX}
\begin{figure*}[t]
\centering
\includegraphics[width=135mm]{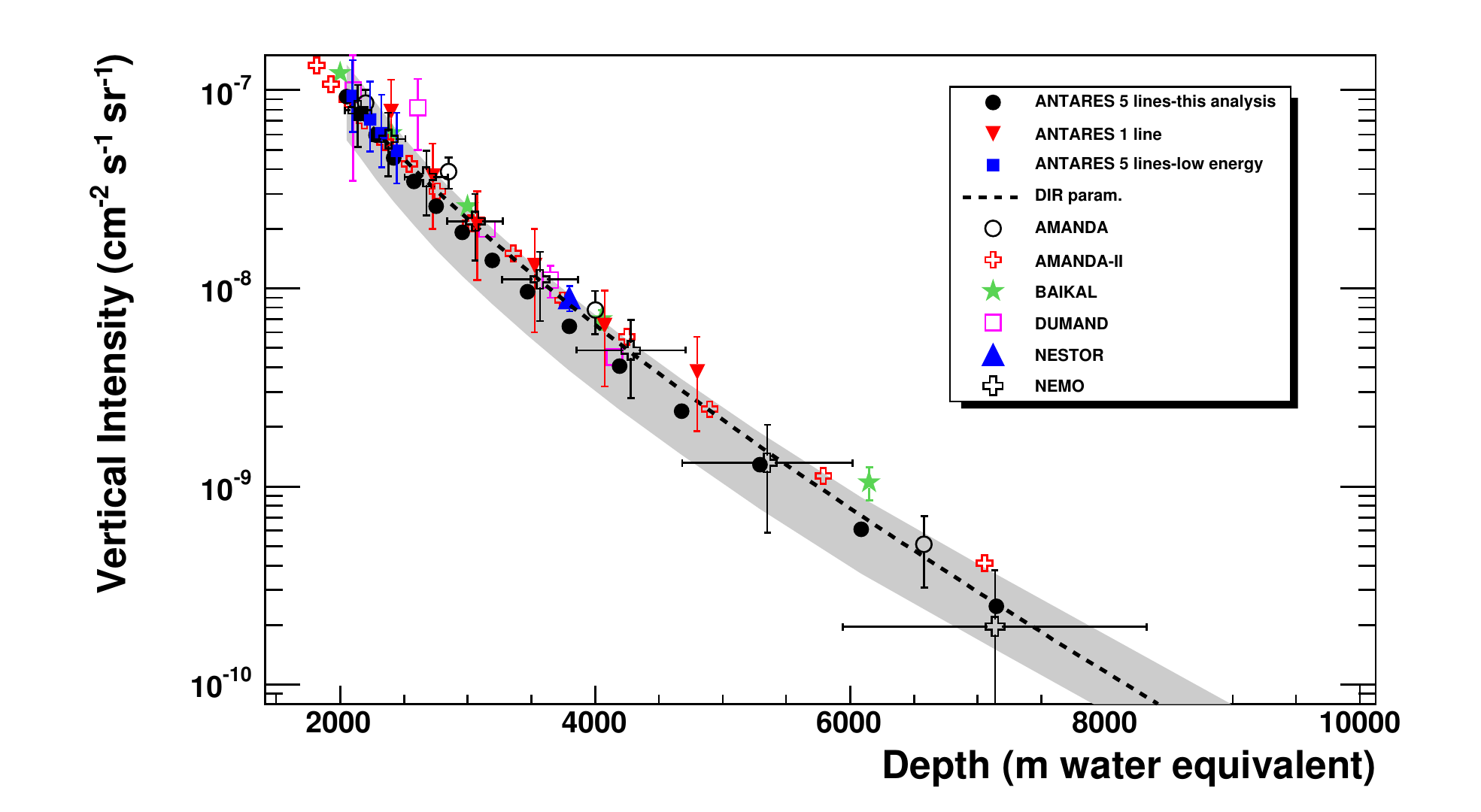}
\caption{Vertical flux of atmospheric muons as a function of the
  equivalent slant depth (taken from \cite{margiotta}), measured with
  5 lines of the ANTARES detector during the construction phase in 2007.} \label{fig:muonflux}
\end{figure*}
The main goal of a neutrino telescope is to detect high-energy neutrinos from
extraterrestrial sources. However, the signal observed by ANTARES is dominated by 
muons that are generated in cosmic ray interactions in the atmosphere above the detector
and which have sufficient energy to reach the detector at its average installation depth
of $2200\,\mathrm{m}$ below sea surface. 
The muon flux measured at the ANTARES site is an important
test beam to study detector systematics and to validate the
reconstruction algorithms employed. Two different studies of the depth-intensity
relation for muons have been carried out. In the first, the attenuation of the muon
flux as a function of depth is observed as a reduction in the rate of photon 
coincidences between adjacent storeys along the detection lines
\cite{zaborov}. This method has the advantage that it does not rely on any 
track reconstruction method and therefore allows for testing directly the response of the 
detector. The second method is based on a standard tracking algorithm that allows for
reconstructing the (average) zenith angle of the incident muon (bundle) which is then used
to compute the track length from the sea surface to the detector.
This track length is usually called ``equivalent slant depth''. 
Taking into
account the known angular distribution of the incident muons, a depth-intensity relation
can be extracted \cite{margiotta}. The results are in good agreement with previous
measurements as can be seen from Fig.~\ref{fig:muonflux}.
 The rather large error band is mainly due
to the systematic uncertainty on the determination of the absorption length of 
light in water and of the angular acceptance of the OMs at large angles,
which becomes important for muons traversing the detector from above due to 
the downward-pointing setup of the OMs.

\section{SEARCH FOR A DIFFUSE NEUTRINO FLUX}
\begin{figure*}[t]
\centering
\includegraphics[width=135mm]{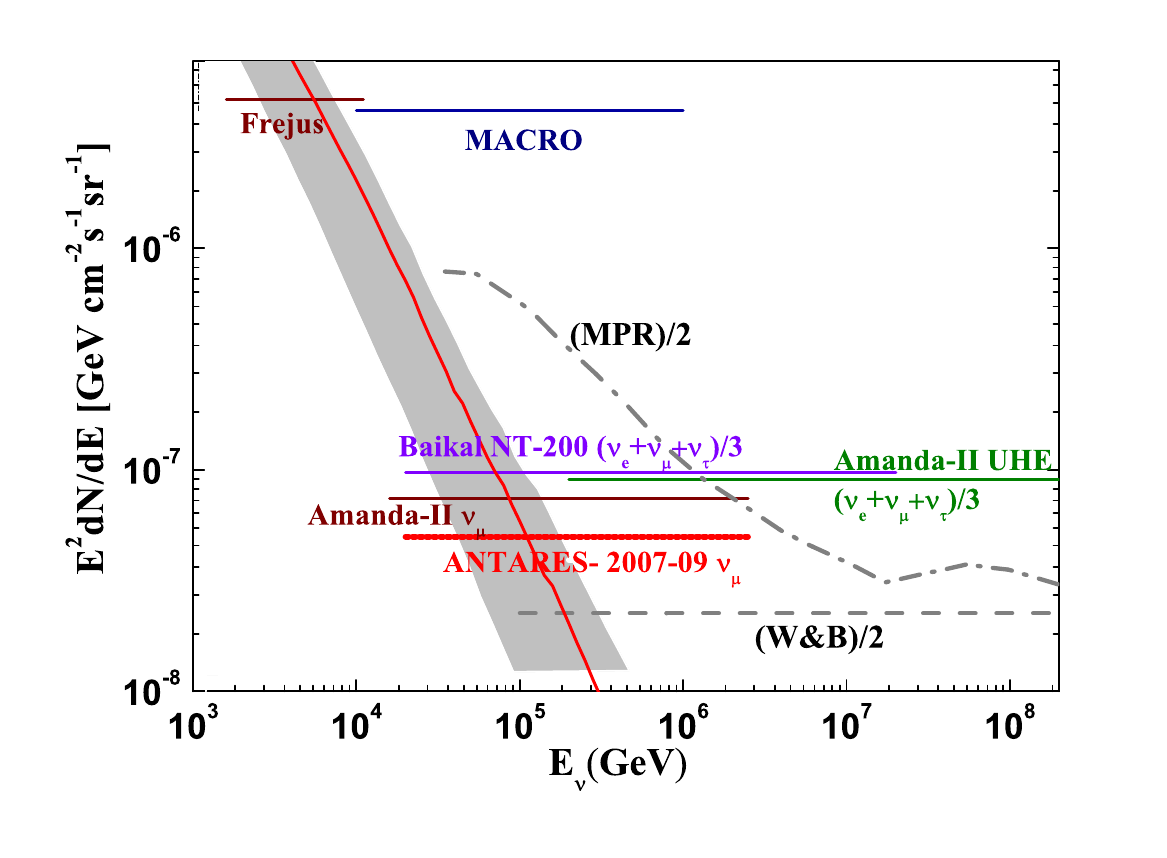}
\caption{The ANTARES upper limit (90\% C.L.) for a $E^{-2}$ diffuse high-energy
$\nu_{\mu} + \overline{\nu}_{\mu}$ flux (taken from \cite{spurio}), 
compared with limits from other experiments and theoretical predictions 
for transparent sources. The factor 1/2 for the W\&B and the MPR model
accounts for neutrino oscillations.
While the central red line represents the average atmospheric neutrino flux,
the grey band denotes the uncertainty due to incident angle and different neutrino
production channels \cite{barr}.} \label{fig:diffuse}
\end{figure*}

The prediction of a neutrino flux from extraterrestrial sources is a direct consequence
of the observation of high-energy particle and gamma radiation impinging on the
Earth's atmosphere \cite{stecker}. While both electrons and charged hadrons can be present at
cosmic acceleration sites, only in the case of hadron acceleration 
will the energy escaping from the source be distributed between the cosmic ray
component, gamma rays and neutrinos. A spatially unresolved, hence diffuse, 
flux of such high-energy neutrinos resulting from different cosmic sources has 
been predicted by various authors. There are two relevant upper bound 
estimates: Waxman and Bahcall (W\&B) \cite{wb1, wb2} 
use as a constraint the cosmic ray flux 
measurements at energies $E_{\mathrm{CR}} \approx 10^{19}\,\mathrm{eV}$; Mannheim, Protheroe
and Rachen (MPR) \cite{mpr} consider the diffuse $\gamma$-ray flux in addition. For sources
that are assumed to be transparent to neutrons, the resulting upper limits are shown
in Fig.~\ref{fig:diffuse}.
The search method for the diffuse neutrino flux exploits the fact that the atmospheric
neutrino flux, which constitutes the main background in this search, has been measured
to exhibit a $E^{-3.7}$ dependence at high energies.
The predicted
diffuse flux of cosmic neutrinos, however, is expected to follow the much harder
energy spectrum of its parent hadron distribution, i.e. a spectrum with a
spectral index close to -2. This prediction results from the fact that the only
known mechanism that can accelerate cosmic ray particles up to the highest observed
energies is the so called Fermi acceleration expected to occur in 
hydrodynamical shock fronts.
To separate atmospheric and diffuse cosmic neutrino fluxes,
a robust energy estimator for high-energy muon neutrino events
has been developed for ANTARES.
The algorithm is based on the average number of hit repetitions
($R$) in the OMs due to the different arrival times of so called direct and delayed 
photons.
The number of hit repetitions for a specific OM in a single event is defined as the number of hits
measured in a time interval up to $500\,\mathrm{ns}$ after and including the first hit
that is used for the muon track reconstruction. The estimator $R$ is calculated 
as the average number of repetitions, dividing the sum of the number of repetitions in the individual OMs
by the number of all OMs that contribute at least one hit selected by the track reconstruction algorithm.
Direct photons reach an OM without being scattered on their way
from their Cherenkov vertex along the muon track, 
whereas scattered Cherenkov photons or photons
induced by electromagnetic showers along the muon track are referred to as delayed with
arrival time differences up to hundreds of nanoseconds with respect to direct photons. 
For high muon energies ($E_{\mu} > 1\,\mathrm{TeV}$) energy loss 
contributions due to radiative processes start to dominate and 
increase linearly with the muon energy, thus leading 
to additional delayed light in the detector due to electromagnetic showers.
This is exploited to select neutrino events and to finally discriminate between
atmospheric background and cosmic neutrinos.
Using a large set of atmospheric muon and
neutrino Monte Carlo events, for which the detector response was fully simulated, the event
selection has been optimised before the signal region was uncovered for the
data. The atmospheric neutrino background was modelled following 
the Bartol flux parametrisation \cite{bartol} 
with an additional high-energy component induced
by the decay of charmed mesons (prompt component)\cite{bugaev}.
The signal neutrino flux $\Phi$ was modelled with a 
$E^{-2}$ spectral shape. The discrimination of signal and background neutrinos is
achieved by a single cut on the hit repetition value $R$ of the above defined energy 
estimator. This cut has been optimised to maximise the detector sensitivity using
Monte-Carlo events only. The expected number of remaining atmospheric neutrino events is
$11\pm 2$. The systematic uncertainties are dominated by the uncertainty on the
atmospheric neutrino flux and the detector acceptance including its dependence on
environmental parameters.
Data taken during the years 2007 to 2009 with an equivalent live time of 334 days and with
several different detector configurations are used for the analysis.
After applying the same cut on the data, 9 neutrino candidate events remain, in full
agreement with the background assumption. This result translates into an upper limit 
(90\% C.L.) for the diffuse cosmic neutrino flux of 
$ E^2\Phi_{\mathrm{90\%}} = 5.3 \times 10^{-8}\,\mathrm{GeV\,cm^{-2}\,s^{-1}\,sr^{-1}}$.
This limit \cite{spurio} holds for an assumed  $E^{-2}$ signal spectrum and in a 
neutrino energy range $20\,\mathrm{TeV} < E_{\nu} < 2.5\,\mathrm{PeV}$. Models yielding
spectral shapes different from the generic $E^{-2}$ have been tested and some of them
could be excluded (cf. \cite{spurio} and ref. therein) at the 90\% confidence level.

\section{NEUTRINO POINT SOURCE SEARCH}
\subsection{Time-integrated search}
\begin{figure*}[t]
\centering
\includegraphics[width=135mm]{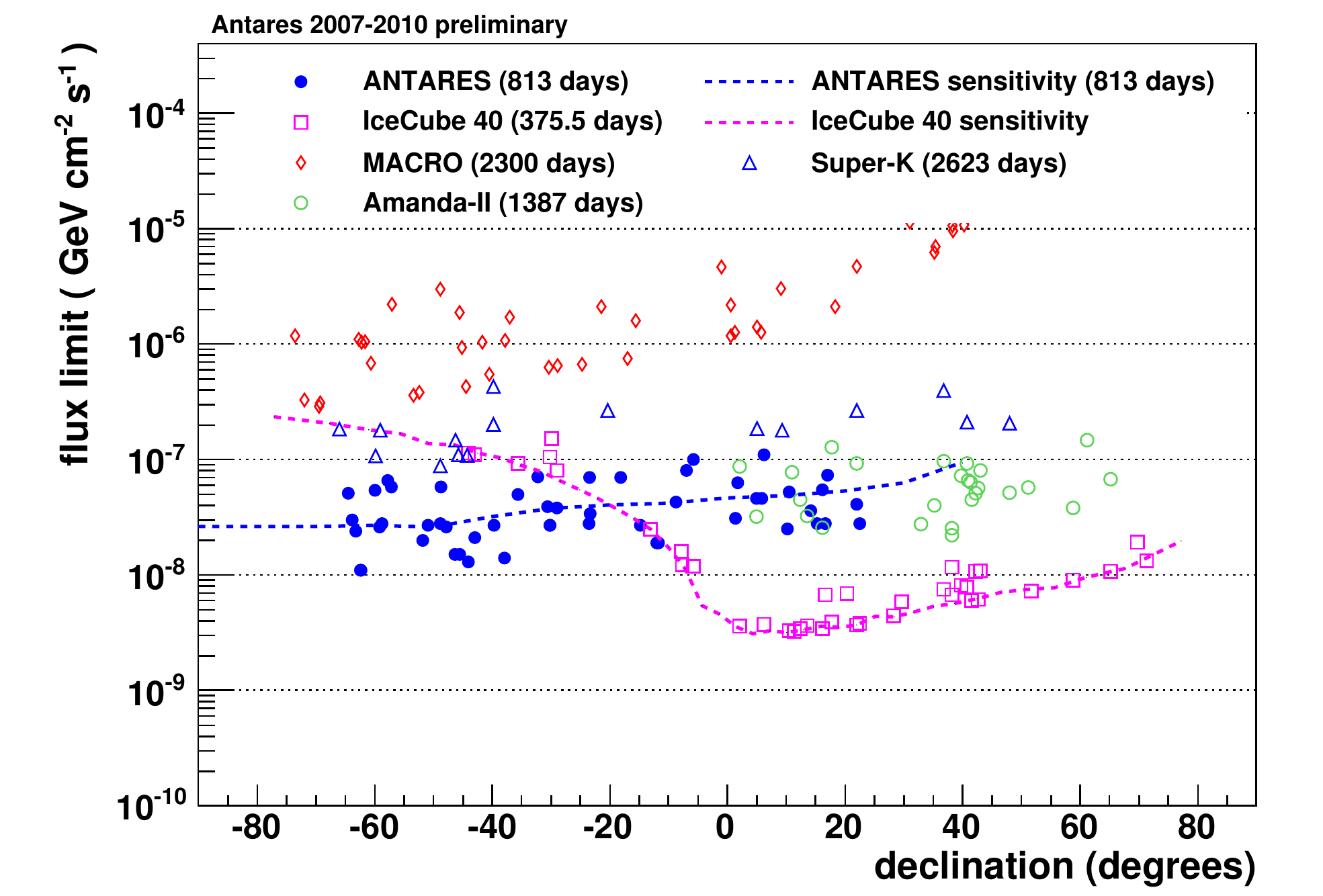}
\caption{Preliminary flux limit (90\% C.L.) vs. source declination
for a list of 50 potential neutrino sources in the 
ANTARES field of view. The sensitivity (blue dashed line) is given as the 
median of the flux limits for the tested sources.} \label{fig:limits}
\end{figure*}
A search for cosmic sources of muon neutrinos has been carried out using data collected
in the years between early 2007 until the end of 2010, corresponding
to an integrated live time of 813 days.
An earlier version of this analysis using data collected in 295
days can be found in \cite{point}. 
The muon reconstruction
algorithm employed is based on a maximum likelihood fit and yields the
track direction, a reconstruction quality parameter based on the reduced
log-likelihood of the track fit and an estimate of the angular error. 
Using a full detector simulation, an average angular resolution
of $0.5^{\circ}\pm 0.1^{\circ}$, defined as the median angle between
the neutrino and the reconstructed muon direction, has been determined
for a $E^{-2}$ neutrino spectrum.
Optimising the limit setting and discovery potential at the same time,
only upward-going events with a good reconstruction quality 
and an angular error estimate better than $1^{\circ}$ are selected. 
The resulting event sample consists of 3058 up-going event candidates.
MC simulations indicate that 84\% of these can be expected to be
neutrinos, while 16\% are expected to be misreconstructed down-going
atmospheric muons. 
As the signature of a point source is a cluster of events at a given celestial position,
two different analyses, both using an unbinned maximum likelihood
method employing the angular resolution and the declination-dependent
rate of background events, are applied to the up-going event sample, in
order to search for such clusters.
In each case, the likelihoods for the assumption that the cluster under
inspection is induced by pure background or by an additional source contribution of
unknown intensity are computed and their ratio is used to distinguish between
signal-like clusters and clusters induced by background fluctuations.
In the first analysis the full sky is searched for point sources,
while in the second analysis neutrinos are searched for only in the
direction of 50 predefined celestial positions, corresponding to
well known galactic and extragalactic astrophysical objects that could
be powerful cosmic accelerators and potential neutrino sources \cite{sourcelist}.
These candidate sources were selected
according to their observed GeV to TeV gamma-ray fluxes, whereby no
temporal coincidence of the gamma-ray emission with the analyzed data 
was required.
Neither search yielded a significant excess of events over the
expectation from pure atmospheric neutrino background.
The post-trial probability of the most signal-like cluster of events
at $(\alpha,\delta) = (-46.5^{\circ},-65.0^{\circ})$ for the full sky
search to be compatible with a fluctuation of the atmospheric neutrino
background is 2.5\% (p-value) and consequently the hypothesis of a neutrino point
source at these coordinates must be rejected as insignificant.
The candidate search yields the object HESS J1023-575 as the
source associated with the most signal-like cluster. The probability,
however, for a background fluctuation resulting in a cluster at least as
significant as the observed one, is 41\%. Consequently, as no excess neutrino
signal has been observed with sufficient significance, upper limits
have been set on 
the flux for each individual source candidate
assuming an $E^{-2}$ spectrum.
The individual limits and the ANTARES sensitivity as a function of
declination and computed as the median
of the limits are reported in  Fig.~\ref{fig:limits}.
The limit computation is based on
a large number of simulated experiments in which systematic
uncertainties on the angular resolution and the acceptance are taken into account.
The obtained upper limits are more stringent than those from previous
experiments in the Northern hemisphere (observing the Southern sky)
and competitive with those set by the IceCube
observatory \cite{icecube-40} for declinations $\delta\lesssim -30^{\circ}$.
The various experiments are sensitive in different energy ranges, even
though they all set limits on $E^{-2}$ spectra. For such a spectrum,
ANTARES detects most events at energies in a broad range around
10 TeV, which is a relevant energy range for several galactic source candidates.

\subsection{Search in coincidence with gamma-ray flaring blazars}
By design, neutrino telescopes constantly monitor at least one
complete hemisphere of the sky and are thus well set to detect
neutrinos produced in transient astrophysical sources.
\begin{figure*}[t]
\centering
\includegraphics[width=135mm]{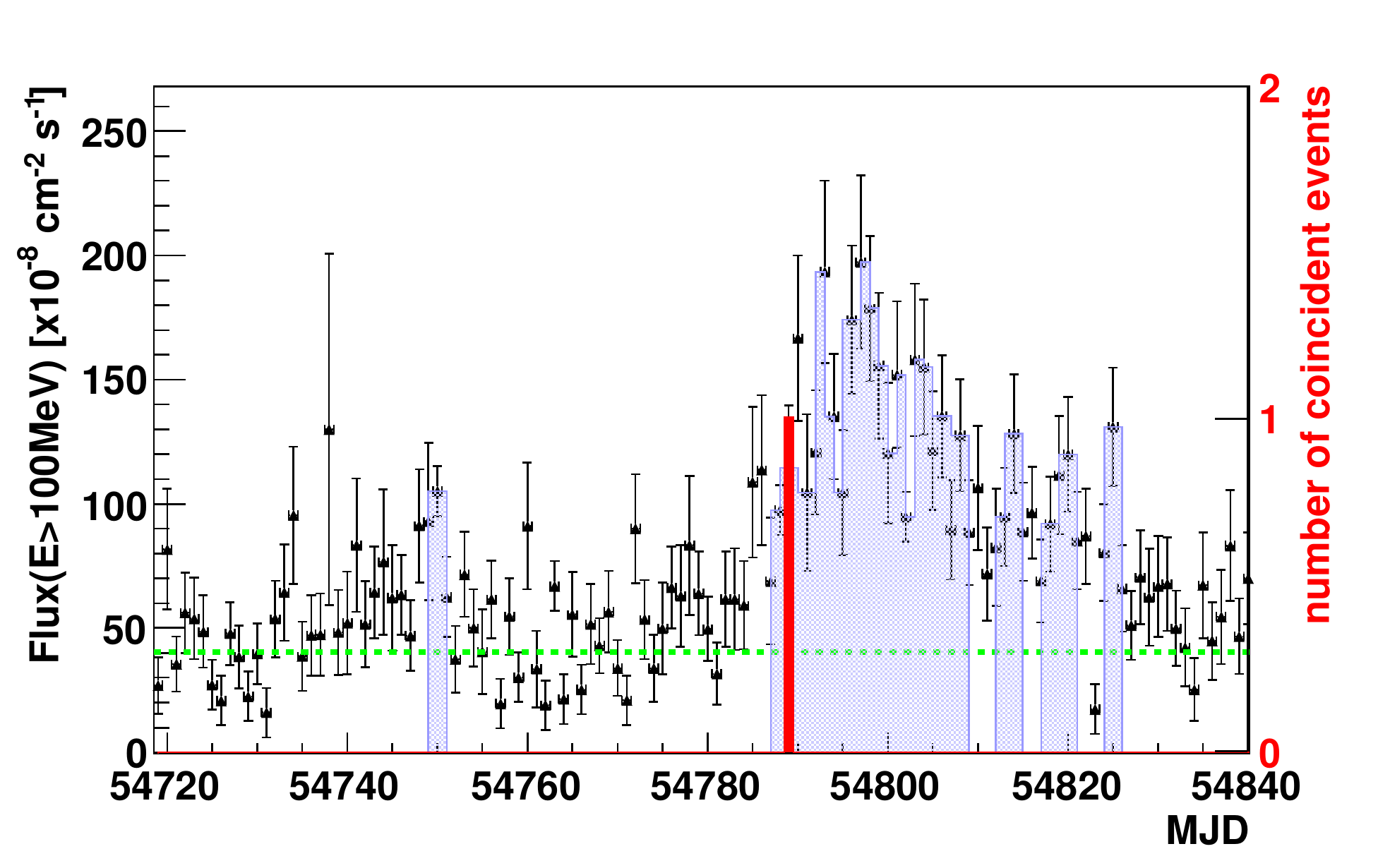}
\caption{Gamma-ray light curve (black data points) of the blazar 3C279
  \cite{3c279} measured by Fermi-LAT with
  $E_{\gamma}>100\,\mathrm{MeV}$. The identified high-state period is
  shown in blue, the baseline fit in green. The time of the coincident
  ANTARES neutrino event is marked in red. Taken from \cite{flares}} \label{fig:fermi}
\end{figure*}
The importance of the irreducible neutrino background
originating from Earth's atmosphere can be drastically reduced
by selecting a narrow time window around the assumed neutrino
production period. Blazars are particularly interesting as
potential neutrino point sources as their enormous energy output
in the form of electromagnetic radiation
and their relativistic outflow of collimated streams of matter
make them good candidate sources of ultra high-energy cosmic rays.
As a consequence, neutrinos and gamma-rays may be
produced in interactions of accelerated hadrons with 
intense ambient photon fields or matter.
The gamma-ray light curves of blazars measured by the LAT instrument
on-board the Fermi satellite reveal important time variability
on a timescale of hours up to several weeks, with high-state intensities mostly several times
larger than the typical flux of the source in its quiescent state \cite{quiescent}.
Assuming a hadronic origin of the observed gamma-rays,
it is assumed in the
following that the observed time-variable gamma-ray fluxes and the  
expected associated neutrino fluxes are proportional. Therefore, high states of
gamma-ray activity in a source are used to define time windows for the
neutrino search from this source. 
The ANTARES data used in this first coincidence search corresponds to the period from
September 6th to December 31st, 2008 and were taken with the complete
detector in its 12-line setup. Periods with high bioluminescence-induced optical
background have been excluded, which results in an effective live time
of 60.8 days. 
The algorithm for a point source search with time dependence
factorizes the probability of a given event to be signal or background into 
a directional and a time component. The probability density function
describing the background contribution is derived from data using the
observed declination distribution of the selected neutrino candidate
events and the time distribution of all reconstructed muons.
The probability density for the signal is described by the telescope's
point spread function and by the gamma-ray light curve of the source
under inspection. Using a method that maximizes the
likelihood ratio of signal and background, the average number of
signal events required to achieve a $5\sigma$ discovery with 50\%
probability can be computed as a function of the flare duration.
For a test flare with a constant flux light curve from a source at a declination
of $\delta=-40^{\circ}$, it has been shown (cf. \cite{flares} for details) that the discovery potential
of the telescope improves at least by a factor of 2 for flare durations
shorter than a day compared to the standard time integrated point
source analysis. 
This time-dependent analysis has been applied to bright and variable
blazars reported in the first year Fermi LAT catalogue \cite{first-fermi-lat-cat}
and in the LAT Bright AGN sample \cite{lbas}. Only sources in the field of
view of ANTARES were selected
whose average high-state photon flux for gamma energies above
$300\,\mathrm{MeV}$
in bins of one day duration
was greater than
$20 \times 10^{-8}\,\mathrm{cm^{-2}\,s^{-1}}$ and 
that showed significant time variability in the studied time period.
The resulting list includes six Flat
Spectrum Radio Quasars and four BL Lacs.
The most significant source is the blazar 3C279,
which has a pre-trial p-value of
1\%. The unbinned method described above finds one high-energy neutrino event
located at $0.6^{\circ}$ from the source location during a large flare in
November 2008. Fig.~\ref{fig:fermi} shows the time distribution of the Fermi
gamma-ray light curve of 3C279 and the time of the coincident neutrino
event \cite{flares}. This event has been reconstructed with 89 hit optical modules
distributed over 10 out of 12 detection lines. The error estimate on
the reconstructed direction, derived from the maximum-likelihood track
fit, is $0.3^{\circ}$. The post-trial probability is computed taking into account the
search for ten different objects. The final probability to
find a signal at least as significant as the one observed amounts to
10\%. Hence, the 
observed neutrino can be attributed to the atmospheric background.
The energy information of the neutrinos has not been used in this
analysis up to now. 

\section{CONCLUSION AND OUTLOOK}
The ANTARES neutrino telescope started routine data taking in a configuration
with 5 installed lines in 2007. Since May 2008 this first-generation telescope is
complete and has in the meanwhile recorded a large neutrino sample of
high quality.
The feasibility of installation and operation of a particle
physics detector in the hostile environment of the deep sea has been demonstrated and
first results have been obtained.
The search for a cosmic diffuse high-energy neutrino flux and a search
for steady point sources both resulted in 
stringent and competitive upper limits for the flux of cosmic neutrinos. 
In order to increase the discovery potential, a first step towards multi-messenger
searches has been made by using Fermi-LAT gamma-ray flux data to
narrow the search time windows for neutrino production. The main
assumption made is that the gamma-ray and neutrino fluxes are
proportional. One neutrino has been found in coincidence with a
gamma-ray flare of a Flat Spectrum Radio Quasar
during the last 4 months of the year 2008, but can be explained
by the atmospheric background. 
The successful operation of the ANTARES neutrino telescope
is an important step towards KM3NeT \cite{km3net},
 a future multi-$\mathrm{km^{3}}$-scale
high-energy neutrino observatory and marine science infrastructure proposed for
construction in the Mediterranean Sea.

\bigskip 
\begin{acknowledgments}
The authors acknowledge the financial support of the funding agencies:
Centre National de la Recherche Scientifique (CNRS), Commissariat \`{a}
l'\'{e}nergie atomique et aux energies alternatives (CEA), Agence National
de la Recherche (ANR), Commission Europe\'{e}nne (FEDER fund and Marie
Curie Program), R\'{e}gion Alsace (contrat CPER), R\'{e}gion
Provence-Alpes-C\^{o}te d'Azur, D\'{e}partement du Var and Ville de La
Seyne-sur-Mer, France; Bundesministerium f\"{u}r Bildung und Forschung
(BMBF), Germany; Istituto Nazionale di Fisica Nucleare (INFN), Italy;
Stichting voor Fundamenteel Onderzoek der Materie (FOM), Nederlandse
organisatie voor Wetenschappelijk Onderzoek (NWO), The Netherlands;
Council of the President of the Russian Federation for young
scientists and leading scientific schools supporting grants, Russia;
National Authority for Scientific Research (ANCS), Romania; Ministerio
de Ciencia e Innovaci\'{o}n (MICINN), Prometeo of Generalitat Valenciana
(GVA) and MultiDark, Spain. We also acknowledge the technical support
of Ifremer, AIM and Foselev Marine for the sea operation and the
CC-IN2P3 for the computing facilities.
\end{acknowledgments}

\bigskip 

\end{document}